\documentclass[12pt]{article}
\usepackage{graphicx}
\begin{document}

\centerline{{\bf Microscopic Abrams-Strogatz model of language
competition}}
\bigskip

\centerline{Dietrich Stauffer*, Xavier Castell\'o, V\'{\i}ctor M. Egu\'{\i}luz, and Maxi San Miguel}

\bigskip
\centerline{IMEDEA (CSIC-UIB), Campus Universitat Illes Balears}

\centerline{E-07122 Palma de Mallorca, Spain}

\bigskip
* Visiting from Institute for Theoretical Physics, Cologne University,

D-50923 K\"oln, Euroland

\bigskip
\centerline{e-mail: \{xavi,maxi,victor\}@imedea.uib.es, stauffer@thp.uni-koeln.de}

\bigskip
{\bf Abstract:}
The differential equations of Abrams and Strogatz for the competition between
two languages are compared with agent-based Monte Carlo simulations for
fully connected networks as well as for lattices in one, two and three
dimensions, with up to $10^9$ agents.


\bigskip
Many computer studies of the competition between different
languages, triggered by Abrams and Strogatz \cite{abrams}, have
appeared mostly in physics journals using differential equations
(mean field approximation \cite{finland,wang,spain,argentina}) or
agent-based simulations for many
\cite{schulze,tibihmo,wichmann,gomes} or few
\cite{kosmidis,schwammle} languages. A longer review is given in
\cite{newbook}, a shorter one in \cite{cise}. We check in this
note to what extent the results of the mean field approximation
are confirmed by agent-based simulations with many individuals. We
do not talk here about the learning of languages
\cite{nowak,roma}.

\begin{figure}[hbt]
\begin{center}
\includegraphics[angle=-90,scale=0.5]{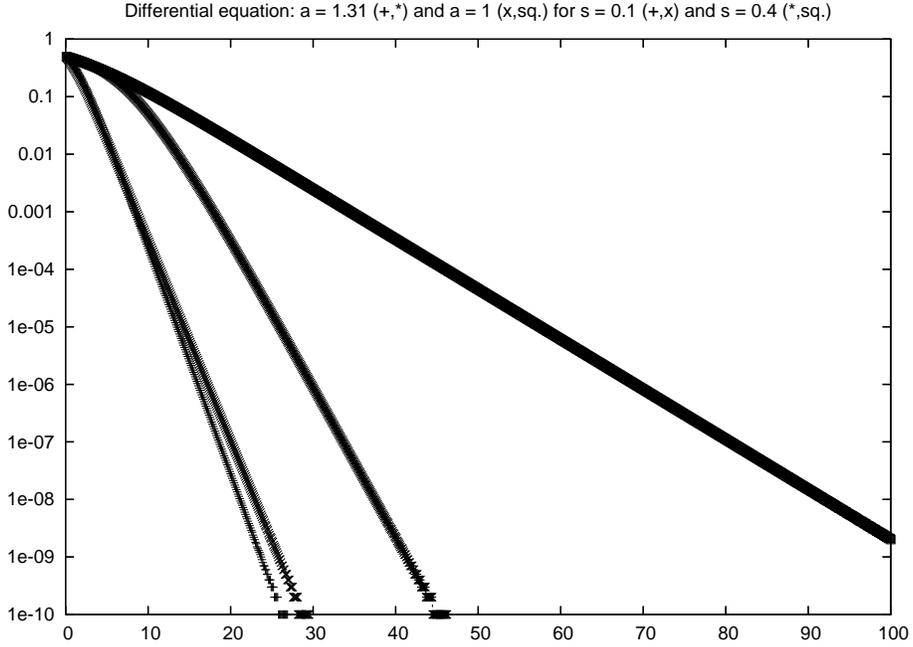}
\end{center}
\caption{
Fraction of X speakers from Abrams-Strogatz differential equation with $a =
1.31$ and $a=1$, at status $s=0.1$ (left) and $s=0.4$ (right). For $a=1.31$ the
decay is faster than for $a=1$.
}
\end{figure}

\begin{figure}[hbt]
\begin{center}
\includegraphics[angle=-90,scale=0.5]{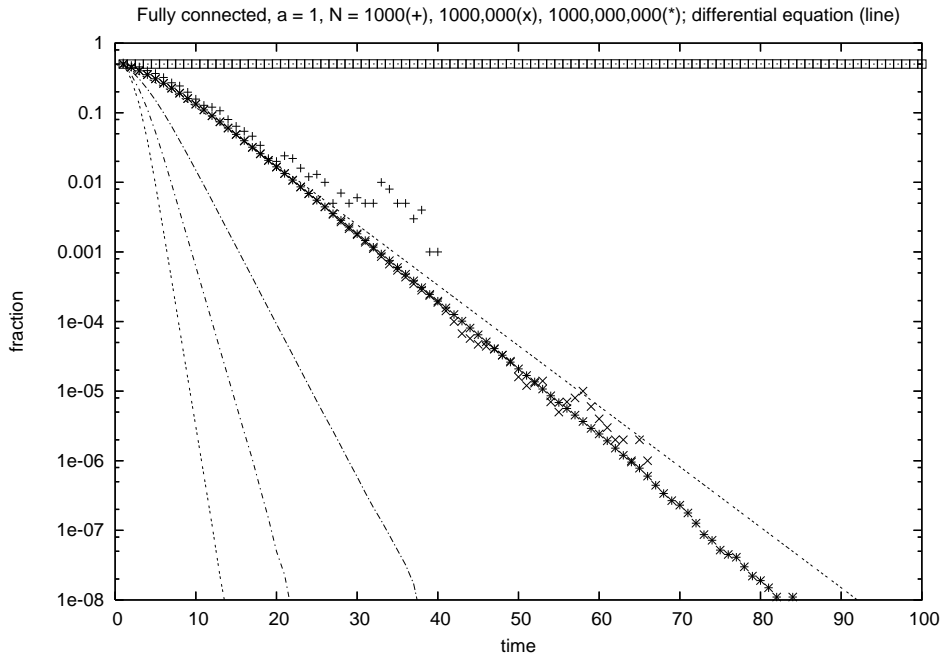}
\end{center}
\caption{ Fully connected model with $10^3, \; 10^6, \; 10^9$
agents at $s = 0.4$ compared with differential equation (rightmost
line) at $s = 0.4$. The three left lines correspond to $s = 0.1,
\; 0.2, \; 0.3$ from left to right for $N = 10^9$. The thick
horizontal line corresponds to $s = 0.5$ and $N = 10^6$ and
changes away from 1/2 only for much longer times. Figs. 2 and 3
use one sample only and thus indicate self-averaging: The
fluctuations decrease for increasing population. }
\end{figure}

\begin{figure}[hbt]
\begin{center}
\includegraphics[angle=-90,scale=0.5]{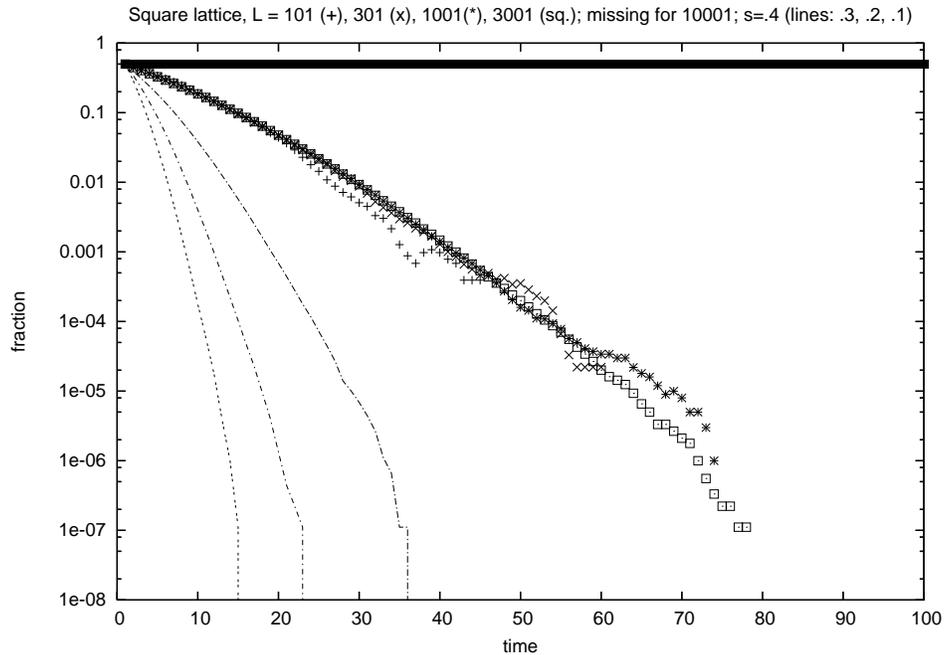}
\end{center}
\caption{ $L \times  L$ square lattice with $L = 101$ to 3001 at
$s = 0.4$. The three left lines correspond to $s = 0.1, \; 0.2, \;
0.3$  from left to right for $L = 3001$. The thick horizontal line
corresponds to $s = 0.5$. }
\end{figure}

\begin{figure}[hbt]
\begin{center}
\includegraphics[angle=-90,scale=0.5]{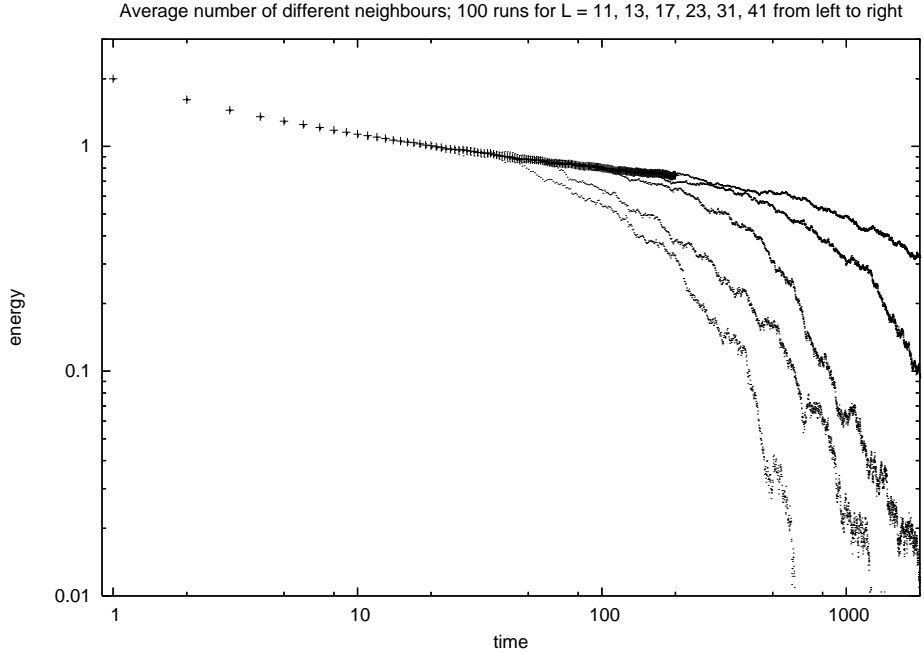}
\end{center}
\caption{ Decay of unstable symmetric solution $x = 1/2$ for
$s=1/2$ for square lattices of various sizes; the larger is the
lattice the longer do we have to wait. A semilogarithmic plot, not
shown, indicates a simple exponential decay. Figs.4-6 average over
100 samples. }
\end{figure}

\begin{figure}[hbt]
\begin{center}
\includegraphics[angle=0,scale=0.40]{abrams5a.eps}
\includegraphics[angle=-90,scale=0.40]{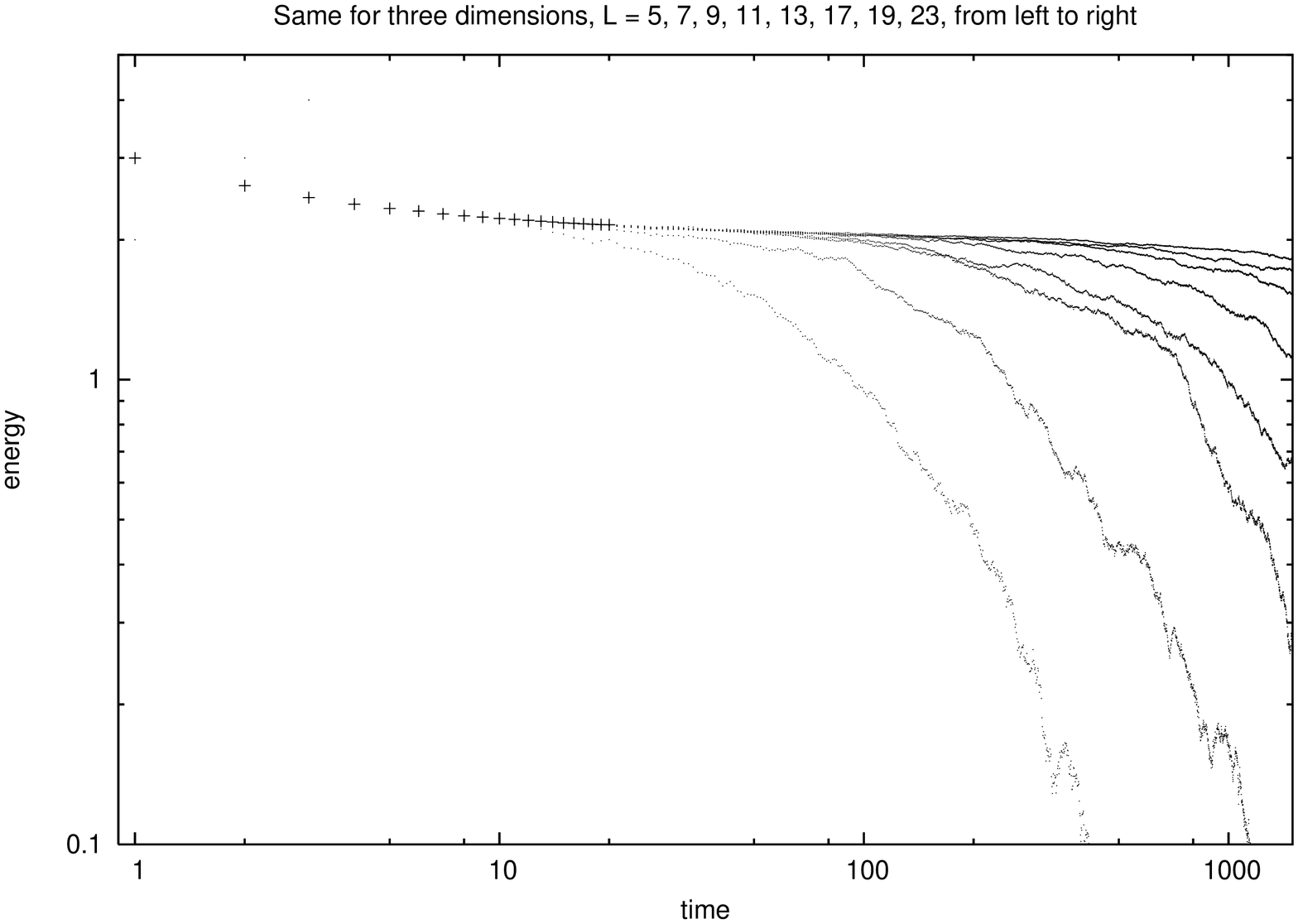}
\end{center}
\caption{ Same as Fig.4 but in one (top) or three (bottom)
dimensions. Simulations shown for $d=1$ are done with random
updating}
\end{figure}

\begin{figure}[hbt]
\begin{center}
\includegraphics[angle=-90,scale=0.40]{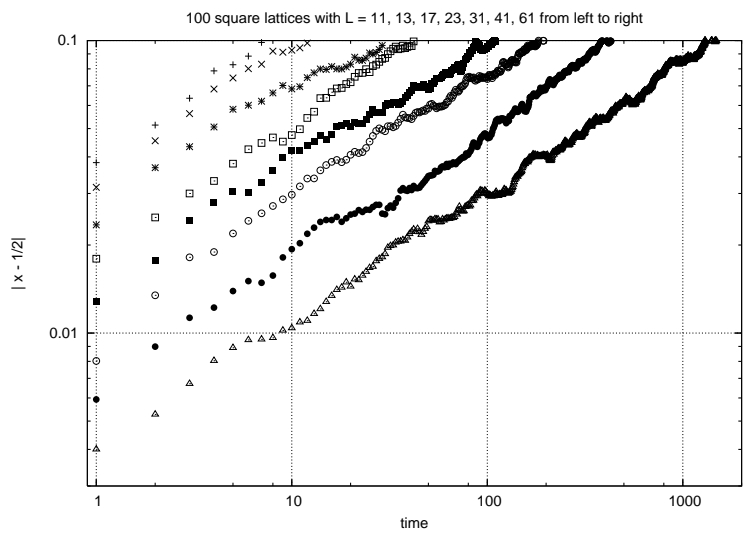}
\end{center}
\caption{ Average over absolute difference between $x(t)$ and
$x(t=0) = 1/2)$ for $d=2$. }
\end{figure}

\begin{figure}[hbt]
\begin{center}
\includegraphics[angle=-90,scale=0.40]{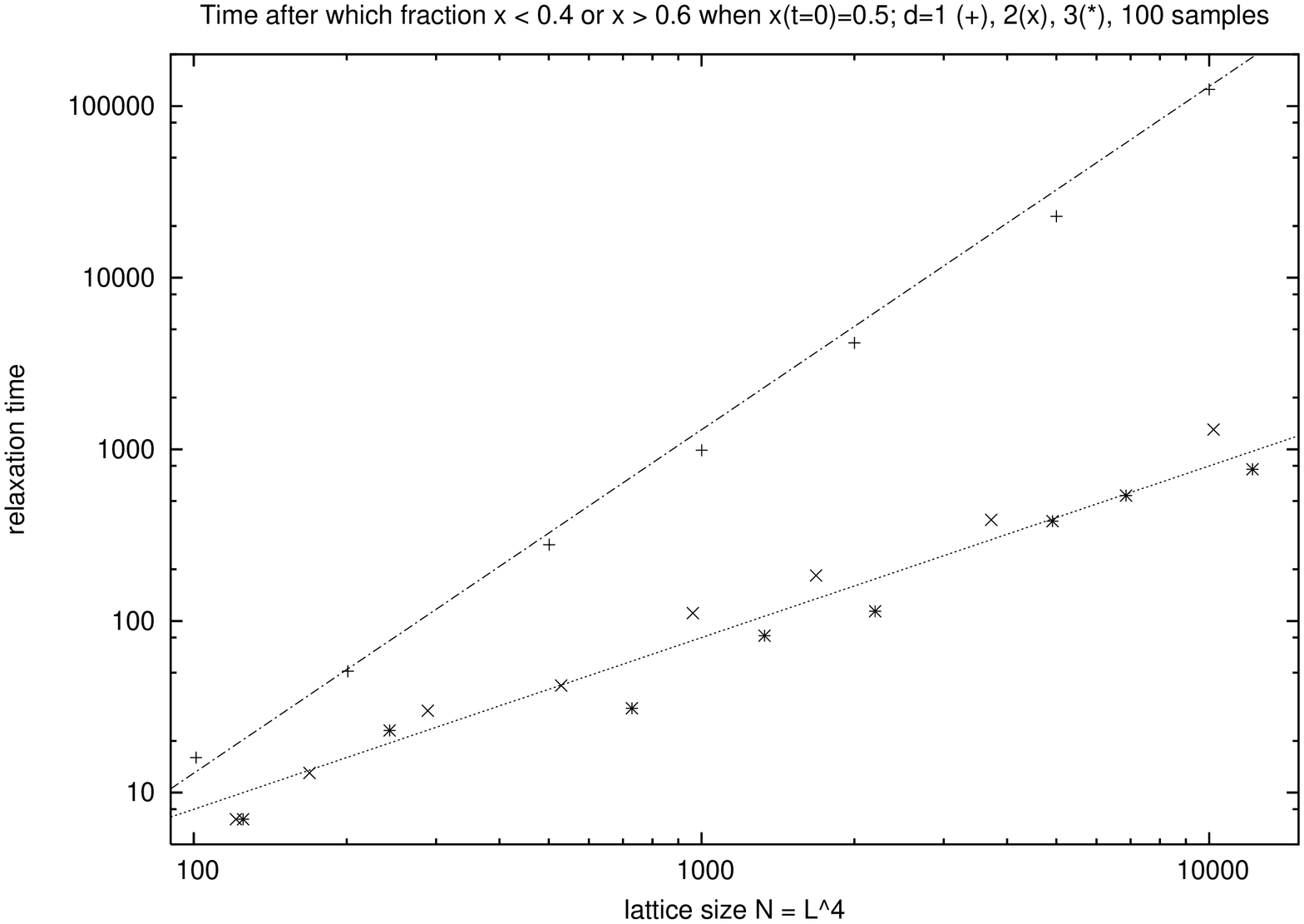}
\includegraphics[angle=0,scale=0.40]{abrams7b.eps}
\end{center}
\caption{ Time for the energy (= number of different lattice
neighbours) to sink to some constant fraction of its initial
value, versus population $N = L^d$, in one (+), two (x) and three
(*) dimensions, from $x(t)$ averaged over 100 samples. Part a uses
regular updating, part b the better random updating. The straight
lines have slope 1 for $d=2,3$, and 2 for $d=1$. }
\end{figure}

The Abrams-Strogatz differential equation for the competition of a language Y
with higher social status $1-s$ against another language X with lower social
status $s$ is
$$ {\rm d}x/{\rm d}t = (1-x)x^as - x(1-x)^a(1-s) \eqno(1) $$
where $a \simeq 1.3$ \cite{abrams} and $0 < s \le 1/2$. Here $x$
is the fraction in the population speaking language X with lower
social status $s$ while the fraction $1-x$ speaks language Y. As
initial condition we consider the situation in which both
languages have the same number of speakers, $x(t=0) = 1/2$. Figure
1 shows exponential decay for $a = 1.31$ as well as for the
simpler linear case $a = 1$. For $s=1/2$ the symmetric situation
$x = 1/2$ is a stationary solution which is stable for $a < 1$ and
unstable for $a > 1$. From now on we use $a=1$. This
simplification makes the resulting differential equation
$$ {\rm d}x/{\rm d}t = (2s-1)(1-x)x \eqno(2) $$
for $s \ne 1/2$ similar to the logistic equation which was applied
to languages before, as reviewed by \cite{ke}. For $s=1/2$ any
value of $x$ is a marginally stable stationary solution.

This differential equation is a mean-field approximation, ignoring
the fate of individuals and the resulting fluctuations. We now put
in $N$ individuals which in the fully connected model feel the
influence of all individuals, while on the $d$-dimensional lattice
they feel only the influence of their $2d$ nearest neighbors. The
probability $p$ to switch from language Y to language X, and the
probability $q$ for the inverse switch, are
$$ p = x^a s, \quad q = (1-x)^a(1-s) \quad. \eqno(3)$$
On a lattice, this $x$ is replaced by the fraction of X speakers
in the neighborhood of $2d$ sites.  We use regular updating for
most of the results shown in this paper. Initially each person
selects randomly one of the two languages with equal probability:
$x(t=0)=0.5$. In the symmetric situation $s=1/2$ with $a=1$ that
we will consider, our later lattice model becomes similar to the
voter model \cite{liggett}.

Fig.2 shows our results for the fully connected case and Fig.3 for
the square lattice with four neighbours; the results are quite
similar to each other and to the original differential equation. A
major difference with the differential equation (1) is seen in the
symmetric case $s = 1/2$ when the two languages are completely
equivalent. Then the differential equation has $x$ staying at 1/2
for all times, while random fluctuation for finite population
destabilize this situation and let one of the two languages win
over the other, with $x$ going to zero or unity.

This latter case can be  described in a unified way by looking at
the number of lattice neighbours speaking a language different
from the center site. It corresponds to an energy in the Ising
magnet and measures microscopic interfaces. Initially this number
equals $d$ on average, and then it decays to zero, first possibly
as a power law, and then exponentially after a time which
increases with increasing lattice size, Fig.4. The first decay
describes a coarsening phenomenon, while the exponential decay is
triggered by finite size fluctuations. In one dimension the
initial decay follows a power law, $t^{-1/2}$, while in three
dimensions an initial plateau is reached. This is followed by an
exponential decay in $d=1,3$ as in two dimensions, Fig.5. Figure 6
shows that the average of $|x(t) - 1/2|$ increases in two
dimensions roughly as the square-root of time until it saturates
at 1/2, indicating random walk behavior. (Note that first
averaging over $x$ and then taking the absolute value $|<x>-1/2|$
would not give appropriate results since $<x>$ would always be 1/2
apart from fluctuations.)

In all the simulations described above, we went through the
population regularly, like a typewriter on a square lattice, and
for full connectivity kept the probabilities constant within each
iteration. Using random updating is more realistic but takes more
time. The long-time results are similar, and the power-law decay
holds for $t < 10^2$ with exponents 0.5 for $d=1$ (Fig. 5), and
0.1 (compatible with $1/\ln t$) for $d=2$. For $d=3$ a plateau is
also reached. For the simpler regular updating we checked when the
fraction $x$, initially 1/2, leaves the interval from 0.4 to 0.6
on its way to zero or one, Fig.7a. For the random updating  we
checked when the energy reaches a small fraction of its initial
value, taken as $2/L$, $0.04$ and $0.6$ for $d=1,2,3$, Fig.7. Both
figure parts are quite similar, with scaling laws for the
characteristic time which are compatible with the ones obtained
for a voter model \cite{liggett}: $\tau \simeq N^2$ in $d=1$,
$\tau \simeq N \ln N$ in $d=2$, and $\tau \simeq N$ in $d=3$,
where $N = L^d$.

We conclude that agent-based simulations differ appreciably from
the results from the mean-field approach for the symmetric case $s
= 1/2$: While Eqs.(1,2) then predict $x$ to stay at $x = 1/2$, our
simulations in Fig.4 and later show that after a decay everybody
speaks the same language. In a fully connected network and in
$d=3$ the decay is triggered by a finite size fluctuation, while
in $d=1,2$ the intrinsic dynamics of the system causes an initial
ordering phenomena in which spatial domains of speakers of the
same language grow in size.

We acknowledge financial support form the MEC(Spain) through
project CONOCE2 (FIS2004-00953).

\end{document}